\documentclass[twocolumn,english,preprintnumbers,amsmath,amssymb,floatfix]{revtex4}
\usepackage[T1]{fontenc}
\usepackage{color}
\usepackage{array}
\usepackage{amstext}
\usepackage{graphicx}
\usepackage{esint}
\usepackage[colorlinks = true,
linkcolor = magenta,
urlcolor  = blue,
citecolor = red,
anchorcolor = blue]{hyperref}

\makeatletter

\@ifundefined{textcolor}{}
{%
	\definecolor{BLACK}{gray}{0}
	\definecolor{WHITE}{gray}{1}
	\definecolor{RED}{rgb}{1,0,0}
	\definecolor{GREEN}{rgb}{0,1,0}
	\definecolor{BLUE}{rgb}{0,0,1}
	\definecolor{CYAN}{cmyk}{1,0,0,0}
	\definecolor{MAGENTA}{cmyk}{0,1,0,0}
	\definecolor{YELLOW}{cmyk}{0,0,1,0}
}

\@ifundefined{definecolor}
{\usepackage{color}}{}
\@ifundefined{definecolor}
{\usepackage{color}}{}
\makeatother

\makeatother

\usepackage{babel}
\begin{document}
	
\title{Probing Heavy Charged Higgs Bosons through Bottom Flavored Hadrons in the $H^+\to \bar{b}t\to B+X$ Channel  in 2HDM}

\author{S. Mohammad Moosavi Nejad$^{a,b}$}
\email{mmoosavi@yazd.ac.ir}
	
\author{ P. Sartipi Yarahmadi$^a$}

	\affiliation{$^{(a)}$Faculty of Physics, Yazd University, P.O. Box
		89195-741, Yazd, Iran\\	
       $^{(b)}$School of Particles and Accelerators,
		Institute for Research in Fundamental Sciences (IPM), P.O.Box
		19395-5531, Tehran, Iran}

	\date{\today}
	
\begin{abstract}

Observing light or heavy charged Higgs bosons $H^\pm$, lighter or heavier than the top quark, would be instant evidence of physics beyond the Standard Model. For this reason, in recent years searches for charged Higgs bosons have been in the center of attention of current colliders such as the CERN Large Hadron Collider (LHC). In spite of all efforts, no signal has been yet observed. Especially, the results of CMS and ATLAS experiments have excluded a large region in the MSSM $m_{H^+}-\tan\beta$ parameter space for $m_{H^+}=80-160$~GeV  corresponding to the entire range of $\tan\beta$ up to 60. Therefore,  it seems that one should concentrate on probing heavy charged Higgs bosons ($m_{H^\pm}>m_t$) so in this context each new probing channel is welcomed.  
In this  work, we intend to present our proposed channel to search for heavy charged Higgses through the  study of scaled-energy distribution of bottom-flavored mesons ($B$) inclusively produced in charged Higgs decay, i.e., $H^+\to t\bar{b}\to B+X$.  Our study is carried out within the framework of the generic two Higgs doublet model (2HDM) using  the massless scheme where the zero mass parton approximation is adopted for bottom quark. 

\end{abstract}

%\pacs{14.65.Ha, 13.88.+e, 14.40.Lb, 14.40.Nd}

\maketitle

\section{Introduction}
\label{sec:intro}

Despite of all successes of the standard model (SM), this model does not represent a theory of everything since there remain many unsolved open questions such as the origin of dark matter, matter-antimatter asymmetry in the universe, the hierarchy problem, etc. To  solve these problems many theories have been proposed which are generally qualified as the theories beyond the SM (BSM). Among the most important ones are those based on the supersymmetry.  These extended models often contain an extended Higgs sector. As an overview, the minimal extensions known as two-Higgs-doublet models (2HDMs)  \cite{Lee:1973iz} include a second complex Higgs doublet which, after spontaneous symmetry breaking, leads to five physical Higgs boson states, i.e., two neutral scalars ($h$ and $H$, with the assumption $m_h < m_H$), two charged Higgs bosons ($H^\pm$) and one neutral pseudoscalar ($A$)  \cite{Djouadi:2005gj}. 
Furthermore, after imposing a discrete symmetry that gives natural flavor conservation the 2HDMs can be  also classified into four categories; Type I, II, III and IV,  according to the couplings of the doublets to the fermions. 
The minimal supersymmetric standard model (MSSM) \cite{Gunion} is one of the most popular and very well-studied BSM scenarios where one doublet couples to up quarks and the other to down quarks and charged leptons. It should be noted that, the Higgs sector of the MSSM is a Type-II 2HDM which provides elegant solutions to some of the short comings of the SM. It does also predict rich and various phenomenology to be testable  in colliders.

Since there is no fundamental charged scalar boson  in the SM, then  the  discovery of a charged scalar boson  would clearly represent unambiguous evidence for the presence of new physics beyond the standard model. In this context, searching for the charged Higgs bosons signal is unique and  in this work we propose a new channel to search for them at the current and future colliders.\\
In  all classes of 2HDM scenario, the charged Higgs bosons $H^\pm$ can appear lighter or heavier than the top quark, while  the lightest CP-even Higgs boson $h$ can align with the properties of the SM. Therefore, looking for charged  Higgs bosons $H^\pm$ in various decay channels over a wide range of masses is a top priority program in the current LHC experiments and future colliders. 

Experimental searches for light charged Higgs bosons ($m_{H^\pm}< m_t$) have already been started at the Tevatron. For example,  the CMS \cite{CMS:2014cdp} and the ATLAS \cite{TheATLAScollaboration:2013wia}  collaborations have reported their results of proton-proton collision data recorded at $\sqrt{s}=8$~TeV using the $\tau+jets$ channel with a hadronically decaying $\tau$ lepton in the final state, i.e., $t\rightarrow  bH^+(\rightarrow \tau^+\nu_\tau)$. Last results on searching for charged Higgs bosons in the $H^\pm\to \tau^\pm\nu_\tau$ decay channel in proton-proton collisions at $\sqrt{s}=13$~TeV is reported by the CMS experiment \cite{Sirunyan:2019hkq}.
According to reported results, the large region in the MSSM $m_{H^+}-\tan\beta$ parameter space is excluded for $m_{H^+}=80-160$~GeV  corresponding to the entire range of $\tan\beta$ up to 60, except a hole around $m_{H^+}\approx 150-160$~GeV for $\tan\beta\approx 10$. Here,  $\tan\beta$ is the ratio of the vacuum expectation values of the neutral components of the two Higgs doublets. Therefore, it seems that there is no much chance to find the light charged Higgs bosons and colliders  should concentrate on probing the heavy charged Higgs bosons ($m_{H^\pm}>m_t$).

Heavy charged Higgs bosons are  mainly produced directly in association with a top quark (and also a bottom quark) \cite{Harlander:2011aa,deFlorian:2016spz}. Moreover,  charged Higgs bosons can be produced in supersymmetric (SUSY) cascade decays via heavier  neutralino and chargino production in squark and  gluino decays, see Refs.~\cite{Datta:2001qs,Datta:2003iz}. 
On the other hand, in many models a heavy charged Higgs boson is predicted to decay predominantly 
either to a tau and its associated neutrino, or to a top and a bottom quark ($H^+\to t\bar{b}$). However, the channel $H^+\to t\bar{b}$ suffers from large multi-jet background, but it dominates in the heavy mass region, see Refs.~\cite{Sirunyan:2020hwv,ATLAS:2020jqj,ATLAS:2016qiq,Aad:2015typ}.
Searches for the signature $H^+\to t\bar{b}$   have been interpreted by the ATLAS and CMS Collaborations in proton-proton collisions at center-of-mass energies of 8 \cite{Aad:2015typ} and 13 TeV \cite{Sirunyan:2020hwv,ATLAS:2020jqj,Aad:2021xzu} and  a small excluded region in the MSSM $m_{H^+}-\tan\beta$ parameter space has been presented. 
For example, the corresponding searches carried out by ATLAS at $\sqrt{s}=13$~TeV and  the integrated luminosity $L = 13.2$ fb$^{-1}$ have  excluded $m_{H^+}\approx 300-900$~GeV for a very low $\tan\beta (\approx 0.5-1.7)$ region \cite{ATLAS:2016qiq}, where as for high values of $\tan\beta >44(60)$, $m_{H^+}\approx 300(366)$~GeV have been  excluded. Therefore, large regions in the parameter space  are still allowed and corresponding searches are in progress.\\
In the present work, we study the dominant decay mode $H^+\to t\bar{b}$ followed by $b\to B+X$, where $B$ is the bottom-flavored hadron and $X$ collectively denotes the unobserved final state particles. Therefore, our proposed channel to search for heavy charged Higgs bosons at colliders is to  study  the energy distributions of B-hadrons  inclusively produced in the decay mode $H^+\to B+X$. 
To this aim, our primary purpose is the evaluation of the next-to-leading order (NLO) QCD corrections to the differential partial decay width $d\Gamma(H^+\to t\bar{b}(+g))/dx_b$, where $x_b$ stands for the scaled-energy  of  bottom quark. This differential  width, which is presented for the first time, is needed to obtain the energy spectrum of B-mesons through heavy charged Higgs decays. Also, the hadronization process $b\to B$ is described by the nonperturbative fragmentation functions (FFs) which will be introduced in Section~\ref{sec:two}. The differential decay width at the parton level ($d\Gamma/dx_b$), the nonperturbative FFs and the factorization theorem, introduced in Sec.~\ref{sec:two}, allow us to compute the desired physical quantity; the energy spectrum of B-hadrons.
Beforehand, in Ref.~\cite{Kniehl:2012mn} we have studied the energy spectrum of B-mesons produced form direct decay of top quarks in the SM, i.e., $t\to BW^++X$. It would be expected that  a comparison between the energy spectrum of B-mesons from charged Higgs decays and those from top decays at SM  indicates  a signal for new physics beyond the SM.
\\
This paper is organized as follows.
In Sec.~\ref{sec:one}, we express  our analytical results of the ${\cal O}(\alpha_s)$ QCD corrections to the Born level rate of $H^+\to t\bar{b}$. We shall apply the massless scheme where the  bottom quark mass is ignored  but the arbitrary value of charged Higgs mass  is retained.
In Sec.~\ref{sec:two}, we give our numerical analysis of inclusive production of B-hadrons from heavy charged Higgs decay considering the factorization theorem and the DGLAP evaluation equations.
 Sec.~\ref{sec:three} is devoted to our summary and conclusions.

\section{Parton level results in the general 2HDM}
\label{sec:one}

Assuming $m_{H^+}>m_t$,  we first study the NLO radiative corrections to the partial decay width 
\begin{eqnarray}\label{born}
H^+\to t\bar{b},
\end{eqnarray}
in the general 2HDM, where  $H_1$ and $H_2$ are the doublets  whose vacuum expectation values (VEV's), i.e., $\textbf{v}_1$  and  $\textbf{v}_2$, give masses to the down and up type quarks, respectively. The squared sum of VEV's is fixed by the Fermi constant $G_F$ as  $\textbf{v}_1^2+\textbf{v}_2^2=(\sqrt{2} G_F)^{-1}=(246~GeV)^2$. However, the ratio of two VEV's is a free parameter and can be characterized by the angle $\beta$ by introducing $\tan\beta=\textbf{v}_2/\textbf{v}_1$. 
A linear combination of the charged components of doublets $H_1$ and $H_2$ does also give the observable charged Higgs $H^\pm$, i.e., $H^\pm=H_2^\pm\cos\beta-H_1^\pm\sin\beta$.\\
In a general 2HDM, tree-level flavor-changing neutral currents (FCNC) can be avoided if one does not couple the same Higgs doublet to up- and down-type quarks simultaneously.
Therefore, for our purpose we need the specific models which  naturally stop these problems by restricting the Higgs coupling.
In this context, there are two possibilities (which are also called two models)  for the two Higgs doublets to couple to the quarks.\\
In the first possibility (or model I), the Higgs doublet $H_1$ couples to all bosons and another doublet $H_2$ couples to all  quarks in the same manner as in the SM.  In this model, the Yukawa couplings between  the top- and the bottom-quark and the charged Higgs  are given by the following Lagrangian \cite{GHK}
\begin{eqnarray}\label{modelfirst}
L_1&=&\frac{g_{_W}}{2\sqrt{2}m_W}V_{tb}\cot\beta \bigg\{H^+\bar{t}\big[m_t(1-\gamma_5)-
\nonumber\\
&&m_b(1+\gamma_5)\big]b\bigg\}+H.c,
\end{eqnarray}
where,  $g_W^2=4\sqrt{2} m_W^2 G_F$ and the CKM matrix element is labeled by  $V_{tb}$.\\
In the second possibility (model II),  the doublet $H_1$ couples only to
the right chiral  down-type quarks while the  $H_2$ couples only to the right chiral up-type quarks. In  this model, the charged Higgs boson couplings to fermions are given by the following Lagrangian
\begin{eqnarray}\label{modelsecond}
L_{2}&=&\frac{g_{_W}}{2\sqrt{2}m_W}V_{tb} \bigg\{H^+\bar{t}\big[m_t\cot\beta(1-\gamma_5)+
\nonumber\\
&&m_b\tan\beta(1+\gamma_5)\big] b\bigg\}+H.c .
\end{eqnarray}
These two models are also known as Type-I and Type-II 2HDM scenarios and, as  mentioned in the Introduction, the  MSSM \cite{Fayet:1974pd,Fayet:1976et,Dimopoulos:1981zb} is a special case of a Type-II 2HDM.

For the process (\ref{born}), considering the interaction Lagrangians (\ref{modelfirst}) and (\ref{modelsecond})  the current density is expressed as $J^{\mu}\propto \psi_b(a+b\gamma_5)\bar{\psi}_t$ so that the coupling factors in two models are given by
\begin{eqnarray}\label{model1}
\textbf{model I}:\quad a&=&\frac{g_{_W}}{2\sqrt{2}m_W}V_{tb}(m_t-m_b)\cot\beta,\nonumber\\
b&=&\frac{g_{_W}}{2\sqrt{2}m_W}V_{tb}(m_t+m_b)\cot\beta,
\end{eqnarray}
and
\begin{eqnarray}\label{model2}
\textbf{model II}:\quad
a&=&\frac{g_{_W}}{2\sqrt{2}m_W}V_{tb}(m_t \cot\beta+m_b\tan\beta),\nonumber\\
b&=&\frac{g_{_W}}{2\sqrt{2}m_W}V_{tb}(m_t \cot\beta-m_b\tan\beta).\nonumber\\
\end{eqnarray}
In next section, we describe the technical detail of our calculation for the ${\cal O}(\alpha_s)$ radiative corrections to the tree-level decay rate of $H^+\to t\bar{b}$ using dimensional regularization to regularize all divergences.

\subsection{Born decay width of $H^+\to t\bar{b}$}

The decay process (\ref{born}) is analyzed in the rest frame of the charged Higgs boson. 
It is straightforward to calculate the Born term contribution to the partial decay rate of the process (\ref{born}) in the 2HDM. According to  the given Lagrangian in Eqs.~(\ref{modelfirst}) and (\ref{modelsecond}), the coupling of the charged-Higgs to the fermions (top and bottom quark in (\ref{born}))  can either be expressed as a superposition of  scalar and pseudoscalar coupling factors or as a combination of right- and left-chiral coupling factors \cite{GHK}. Therefore, the lowest order decay amplitude is of the form  
\begin{eqnarray}\label{b1}
M_0=v_b(a\boldsymbol{1}+b\gamma_5)\bar{u}_t=v_b\{g_t\frac{1+\gamma_5}{2}+g_b\frac{1-\gamma_5}{2}\}\bar{u}_t,
\end{eqnarray}
where, $g_t=a+b$ and $g_b=a-b$. 
Therefore, the tree-level decay width reads
\begin{eqnarray}\label{gammatree}
\Gamma_0&=&\frac{N_c m_H}{8\pi}\lambda^{\frac{1}{2}}(1,R,y)\bigg[2(a^2+b^2)(S-R)\nonumber\\
&&-2(a^2-b^2)\sqrt{Ry}\bigg],
\end{eqnarray}
where $\lambda(x,y,z)=(x-y-z)^2-4y z$ is the  K\"all\'en function and $N_c=3$ is a color factor. Here, for simplicity, we have defined: $R=(m_b/m_H)^2$, $y=(m_t/m_H)^2$ and $S=(1+R-y)/2$.
This result is in complete agreement with the one presented in Ref.~\cite{Li:1990ag}.
In the limit of vanishing bottom quark mass, the tree-level decay width is of the form
\begin{eqnarray}\label{gammatreezero}
\Gamma_0=\frac{N_c m_H(1-y)^2}{8\pi}(a^2+b^2),
\end{eqnarray}
where, in both models I and II one has
\begin{eqnarray}\label{gammaaa}
a^2+b^2=\sqrt{2}G_F |V_{tb}|^2 m_t^2\cot^2\beta.
 \end{eqnarray}
Since $m_b<<m_t$, finite-$m_b$ corrections are expected to be negligible in the case at hand. This expectation has been actually confirmed in Ref.~\cite{Kniehl:2012mn}, by a comparative analysis of the partial width of the decay $t\to bW^+$ in the  general-mass variable-flavor-number scheme (GM-VFNS), where bottom-quark mass is preserved, and the zero-mass variable-flavor-number scheme (ZM-VFNS), where bottom is included among the massless quark flavors. Then, throughout this work  we apply the ZM-VFNS or massless scheme.  

In next section,  we compute the ${\cal O}(\alpha_s)$ QCD corrections to the Born-level decay rate of $H^+\to t\bar{b}$ and
present, for the first time, the analytical parton-level expressions for $d\Gamma(H^+\to B+X)/dx_B$ at NLO in the ZM-VFNS. To this aim, we calculate the quantity $d\Gamma_b/dx_b$ where,
\begin{eqnarray}\label{variable}
x_b=\frac{E_b}{E_b^{max}}=\frac{2E_b}{m_H(1-y)},
\end{eqnarray}
is the scaled-energy  of b-quark. It ranges as $0\leq x_b\leq 1$.

\subsection{${\cal O}(\alpha_s)$ virtual corrections}
\label{virtual}

The QCD virtual one-loop  corrections to the process $H^+\to t\bar{b}$  contain  both infrared (IR) and
ultraviolet (UV) divergences where the UV-divergences appear when the integration
region of the internal momentum of the virtual gluon goes to infinity and the
IR-divergences arise from the soft-gluon singularities.
In this work,  we adopt the "on-shell" mass renormalization
scheme and apply the dimensional regularization scheme to regularize all divergences. Through this scheme, all singularities are regularized in $D =4-2\epsilon$   dimensions to become single poles in $\epsilon$.
Considering the two-body phase space for the virtual corrections the contribution of virtual radiations into the
differential decay width reads
\begin{eqnarray}
\frac{d\Gamma^{\textbf{vir}}_b}{dx_b}=\frac{S}{8\pi m_H}
\overline{|M^{\textbf{vir}}|^2}\delta(1-x_b),
\end{eqnarray}
where,
$\overline{|M^{\textbf{vir}}|^2}=\sum_{Spin}(M_0^{\dagger} M_{loop}+M_{loop}^{\dagger} M_0)$. Here, $M_0$ is the Born term amplitude (\ref{b1}) and the renormalized amplitude of the virtual corrections  is given by
$M_{loop}=v_b(\Lambda_{ct}+\Lambda_l)(a+b\gamma_5)\bar{u}_t$,
where  $\Lambda_{ct}$ represents the counterterm and $\Lambda_l$
arises from the one-loop vertex correction \cite{MoosaviNejad:2012ju}.
Following Refs.~\cite{Czarnecki,Liud},  the counterterm
of the vertex includes  the wave-function renormalizations of  quarks  as well as the top quark mass renormalization
\begin{eqnarray}
\Lambda_{ct}=\frac{\delta Z_b}{2}+\frac{\delta Z_t}{2}-
\frac{\delta m_t}{m_t}.
\end{eqnarray}
Since, we are working in the ZM-VFN scheme where $m_b=0$ is assumed, then the b-quark mass counterterm is $\delta m_b=0$.
The  wave function and the mass renormalization constants are given by \cite{Korner:2002fx}
\begin{eqnarray}\label{mass}
\delta Z_b &=& -\frac{\alpha_s(\mu_R)}{4\pi}C_F\big[\frac{1}{\epsilon_{UV}}-\frac{1}{\epsilon_{IR}}\big],
	\nonumber\\
\delta Z_t &=& -\frac{\alpha_s(\mu_R)}{4\pi}C_F\big[\frac{1}{\epsilon_{UV}}+\frac{2}{\epsilon_{IR}}
-3\gamma_E+3\ln\frac{4\pi\mu_F^2}{m_t^2}+4\big],
\nonumber\\
\frac{\delta m_t}{m_t}&=&\frac{\alpha_s(\mu_R)}{4\pi}C_F\big[\frac{3}{\epsilon_{UV}}-3\gamma_E+
3\ln\frac{4\pi\mu_F^2}{m_t^2}+4\big],
\end{eqnarray}
where,  $\gamma_E=0.5772\cdots$ is the Euler constant,   $C_F=(N_c^2-1)/(2N_c)=4/3$ for $N_c=3$ quark colors, and $\mu_F$ is the factorization scale which is arbitrarily set as $\mu_F=m_H$ in our work.
Conventionally, $\epsilon_{IR}$ and $\epsilon_{UV}$ represent the infrared and
the ultraviolet divergences, respectively.\\
The real part of the vertex correction is given by
\begin{eqnarray}
\Lambda_l&=&\frac{\alpha_s N_c m_H^2}{\pi}C_F(a^2+b^2)\big[y-1+(1-y)B_0(0,0,0)\nonumber\\
&&-yB_0(m_H^2,0,m_t^2)+B_0(m_t^2,0,m_t^2)
\nonumber\\
&&-(1-y)^2m_H^2 C_0(0,m_t^2,m_H^2,0,0,m_t^2)\big],
\end{eqnarray}
where, $B_0$ and $C_0$ are the Passarino-Veltman 2-point and 3-point integrals \cite{Dittmaier:2003bc}.
By summing all virtual corrections up, the UV-singularities  are canceled so that the virtual differential decay rate is ultraviolet finite.  But, the IR-divergences are remaining which are now labeled by $\epsilon$.
Eventually, the virtual one-loop contributions read 
\begin{eqnarray}\label{virt}
\frac{d\Gamma^{\textbf{vir}}_b}{dx_b}&=&\Gamma_0
\frac{\alpha_s(\mu_R)}{2\pi}C_F
\delta(1-x_b)\bigg\{2Li_2(y)-\frac{1}{\epsilon^2}+\frac{F}{\epsilon}
\nonumber\\
&&-\frac{F^2}{2}+(2y-5)\ln\frac{1-y}{y}+\ln^2y-\frac{3\pi^2}{4}-\frac{7}{8}\bigg\},
\nonumber\\
\end{eqnarray}
where, $Li_2(y)=-\int_0^y (dt/t) \ln(1-t)$ is the Spence function and
\begin{eqnarray}
F=-\ln\frac{4\pi}{y}+2\ln\frac{1-y}{y}+\gamma_E-\frac{5}{2}.
\end{eqnarray}

\subsection{Real gluon corrections (Bremsstrahlung)}

To obtain the infrared-finite physical results  for $d\Gamma_b/dx_b$  one must include the contributions of real gluons emission. 
Considering two Feynman graphs including the real gluon emissions from the top and bottom quark, the ${\cal O}(\alpha_s)$ real gluon emission (tree-graph) amplitude reads
\begin{eqnarray}\label{finfin}
M^{\textbf{real}}&=&g_s\frac{\lambda^a}{2} v(p_b, s_b)\big\{-\frac{2p_t^\mu+
\displaystyle{\not}p_g \gamma^\mu}{2p_t \cdot p_g}
\\
&&+\frac{p_b^\mu+\gamma^\mu \displaystyle{\not}p_g}
{2p_b\cdot p_g}\big\}(a\textbf{1}+b\gamma_5) \bar{u}(p_t, s_t)\epsilon_{\mu}^{\star}(p_g,r),\nonumber
\end{eqnarray}
where $\epsilon(p_g,r)$ refers to the polarization vector of the  emitted real gluon with the spin $r$. The first and second expressions in the curly brackets are related to the real gluon emissions from the top  and bottom quarks, respectively.
In order to regulate the IR-divergences which arise from the soft and collinear real-gluon emissions, as before, we apply dimensional regularization scheme. According to this scheme, the real differential decay rate for the process $H^+\to t\bar{b}g$ is given by
\begin{eqnarray}\label{moozmooz}
d\Gamma^{\textbf{real}}=\frac{\mu_F^{2(4-D)}}{2m_H}|M^{\textbf{real}}|^2dR_3(p_t, p_b, p_g, p_{_{H^+}}),
\end{eqnarray}
where, $\mu_F$ is an arbitrary reference mass and the phase space element $dR_3$ is defined as
\begin{eqnarray}\label{ahah}
\frac{d^{D-1}\vec{p}_b}{2E_b}\frac{d^{D-1}\vec{p}_t}{2E_t}\frac{d^{D-1}\vec{p}_g}{2E_g}
(2\pi)^{3-2D}\delta^D(p_H-\sum_{g,b,t} p_f).
\end{eqnarray}
To evaluate the differential decay rate $d\Gamma^{real}_b/dx_b$, 
we fix the momentum of bottom quark in Eq.~(\ref{moozmooz}) and integrate over 
the gluon energy which ranges as
\begin{eqnarray}
m_H\frac{(1-y)(1-x_b)}{2}\leq E_g \leq  m_H\frac{(1-y)(1-x_b)}{2(1-x_b(1-y))}.
\end{eqnarray}
Note that, when we integrate over the phase 
space of the real gluon radiation, terms of the form $(1-x_b)^{-1-2\epsilon}$ appear which are due to the radiation of  soft gluon, i.e., $E_g\to 0 \equiv x_b\to1$.  Thus, we employ the following prescription introduced in Ref.~\cite{Corcella:1}
\begin{eqnarray}
(1-x_b)^{-1-2\epsilon}&&=-\frac{1}{2\epsilon}\delta(1-x_b)+\bigg(\frac{1}{1-x_b}\bigg)_+\nonumber\\
&&-2\epsilon \bigg(\frac{\ln(1-x_b)}{1-x_b}\bigg)_+,
\end{eqnarray}
where the plus distributions are  defined as 
\begin{eqnarray}
\int_0^1 (f(x))_{_+}h(x)dx=\int_0^1f(x)[h(x)-h(1)]dx.
\end{eqnarray}

\subsection{Analytical results for $d\Gamma/dx_i$ at parton level}

The ${\cal O}(\alpha_s)$ corrections to the differential decay rate of $H^+\to t\bar{b}$ is obtained by  summing the Born, the virtual and the real gluon contributions.
It reads
\begin{eqnarray}\label{pol1}
\frac{d\Gamma^{\textbf{nlo}}_b}{dx_b}&=&\Gamma_0\Big[\delta(1-x_b)+
\frac{C_F\alpha_s}{2\pi}\Big\{[-\frac{1}{\epsilon}+\gamma_E-\ln 4\pi]\nonumber\\
&&\times[\frac{3}{2}\delta(1-x_b)+\frac{1+x_b^2}{(1-x_b)_+}]+T_1\Big\}\Big],
\end{eqnarray}
where, by defining $S=(1-y)/2$ (with $y=m_t^2/m_H^2$) one has
\begin{eqnarray}\label{pol11}
T_1&=&\delta(1-x_b)\bigg\{\frac{3}{2}\ln y+4S\ln\frac{y}{1-y}-2Li_2\frac{1}{y}-\frac{\pi^2}{3}-2\bigg\}\nonumber\\
&&+2(1+x_b^2)\bigg(\frac{\ln(1-x_b)}{1-x_b}\bigg)_+\nonumber\\
&&+\frac{1+x_b^2}{(1-x_b)_+}\bigg\{\ln\frac{4S^2x_b^2}{1-2Sx_b}+\frac{1}{(1-2Sx_b)^2}\bigg[-2S^2x_b^2\nonumber\\
&&+\frac{(1-x_b)^2+2x_b(4Sx_b-1)}{1+x_b^2}\bigg]\bigg\}.
\end{eqnarray}
Our result of differential decay rate, which is presented for the first time,  after integration over $x_b$ ($0\leq x_b \leq 1$) is in complete agreement with the result presented in \cite{Li:1990ag}.

Note that, our main purpose is to evaluate the energy distribution of B-hadrons produced in heavy charged Higgs boson decay: $H^+\to t\bar{b}(+g)\to B+X$, where B-hadrons can be produced from the fragmentation of b-quark as well as the emitted real gluons. Therefore, in order to obtain 
the most accurate energy spectrum of produced B-hadrons we have to consider  the contribution of gluon fragmentation as well.
It should be noted that,  the gluon splitting contribution is  
important at the  low energy of the observed B-hadron  so this contribution decreases the size of decay rate at the threshold, see Refs.~\cite{Nejad:2016epx,Nejad:2014sla}. With this explanation, we  also need to compute the NLO differential decay rate $d\Gamma^{\textbf{nlo}}_g/dx_g$, where $x_g=2E_g/(m_H (1-y))$ is the scaled-energy of  emitted real gluon, as  in (\ref{variable}).
Ignoring the details of calculation, this differential decay rate is given by
\begin{eqnarray}\label{pol2}
\frac{d\Gamma^{\textbf{nlo}}_g}{dx_g}&=&\\
&&\hspace{-1cm}\Gamma_0\frac{C_F\alpha_s}{2\pi}\bigg\{\frac{1+(1-x_g)^2}{x_g}(-\frac{1}{\epsilon}+\gamma_E-\ln 4\pi)+T_2\bigg\},\nonumber
\end{eqnarray}
where,
\begin{eqnarray}\label{pol22}
T_2&=&\frac{1+(1-x_g)^2}{x_g}\ln\frac{S^2x_g^2(1-x_g)^2(1-2Sx_g)}{y^2}\nonumber\\
&&+\frac{(x_g+2)^2-8}{x_g}.
\end{eqnarray}
In Eqs.~(\ref{pol1}) and (\ref{pol2}), the terms $T_1$ and $T_2$ are free of all IR-divergences. In order to subtract the singularities 
remaining in the differential decay widths, we employ the modified minimal-subtraction $(\overline{MS})$ scheme, where the singularities are absorbed into the bare fragmentation functions (FFs). This renormalizes the
FFs, endowing them with $\mu_F$ dependence, and creates in
the differential decay widths the finite terms of the form $(\alpha_s/\pi)\ln(m_H^2/\mu_F^2)$  which are rendered perturbatively small by choosing $\mu_F={\cal O}(m_H)$.
Following the  $\overline{MS}$ scheme, in order to have  the finite coefficient functions we have to subtract from
Eqs.~(\ref{pol1}) and (\ref{pol2}),  the ${\cal O}(\alpha_s)$ term multiplying the  characteristic $\overline{MS}$ constant, i.e., $-1/\epsilon+\gamma_E-\ln 4\pi$ \cite{Corcella:1}.

\section{Numerical results}
\label{sec:two}

In this work, using the ZM-VFNS we study the decay process 
\begin{eqnarray}\label{eqq}
H^+\to t\bar{b}(+g),
\end{eqnarray}
followed by $\bar{b}/g\to B+X$. In this process,  top quark dominantly decays  as: $t\to bW^+\to bl^+\nu_l$. 
In the narrow-width approximation (NWA), where we set  $p_t^2=m_t^2$ and $p_{W^+}^2=m_{W^+}^2$ and ignore small terms of order ${\cal O}(\Gamma_i^2/m_i^2)(i=t,W^+)$, the total decay rate reads  
\begin{eqnarray}\label{br}
&&\Gamma(H^+\to b\bar{b}l^+\nu_l)=\\
&&\Gamma(H^+\to t\bar{b})\times B(t\to bW^+)\times B(W^+\to l^+\nu_l),\nonumber
\end{eqnarray}
where, for the branching ratios one has $B(t\to bW^+)=96.2\%$ and $B(W^+\to l^+\nu_l)=10.86\%$ \cite{Zyla:2020zbs}.
More details  about the NWA can be found in Ref.~\cite{MoosaviNejad:2019agw}.

Having the differential decay widths  for the process (\ref{eqq}), i.e., Eqs.~(\ref{pol1}) and (\ref{pol2}), we are now in a situation to make our  phenomenological predictions  
for the scaled-energy ($x_B$) distribution of B-hadrons inclusively  produced in the decay of heavy charged Higgs bosons. To present our results for the $x_B$-distribution, we
consider the  differential distribution  $d\Gamma^{nlo}/dx_B$ of the partial width of the decay $H^+\to B+X$, where   
$x_B=2E_B/(m_H(1-y))$ is the scaled-energy of B-hadrons in the charged Higgs rest frame. The $x_B$-variable  is defined as $x_b$  in (\ref{variable}).

Our tool to compute the scaled energy
distribution of  B-hadrons is the factorization theorem of QCD-improved parton model \cite{collins}. According to this theorem \cite{Salajegheh:2018hfs}, the  energy distribution of B-hadrons can be expressed as the convolution of the  parton-level spectrum $d\Gamma_a/dx_a (a=b, g)$ with
the nonperturbative FFs of $a\to B$, describing  the hadronization process of $a\to B$. The $a\to B$ FFs are labeled by  $D_a^B(z, \mu_F)$, where $\mu_F$ is the factorization scale and $z=E_B/E_a$ is the fragmentation  variable which indicates the energy fraction of parent parton carried by the produced hadron. The factorization theorem is expressed as 
\begin{equation}\label{eq:master}
\frac{d\Gamma}{dx_B}=\sum_{a=b, g}\int_{x_a^\text{min}}^{x_a^\text{max}}
\frac{dx_a}{x_a}\,\frac{d\Gamma_a}{dx_a}(\mu_R, \mu_F) D_a^B(\frac{x_B}{x_a}, \mu_F),
\end{equation}
where,  $\mu_R$ and $\mu_F$ are the renormalization and factorization  scales, respectively. The scale $\mu_R$ is  related to the renormalization of the QCD coupling constant. In  this paper, we use the convention $\mu_R=\mu_F=m_H$, a choice often made.

Several searches for the signature $H^+\to t\bar{b}$ in the context of 2HDMs  have been done by the ATLAS and CMS Collaborations in proton-proton  collisions at center-of-mass energies of 8 and 13 TeV \cite{Aad:2015typ,Sirunyan:2020hwv,ATLAS:2020jqj}. For example, in Ref.~\cite{Sirunyan:2020hwv} the presented results  are based on proton-proton collision data collected in 2016 at $\sqrt{s}=13$~TeV by the CMS experiment, corresponding to an integrated luminosity of $35.9~ fb^{-1}$.  Figure 7 in this reference  shows the excluded parameter space in the MSSM scenarios. Based on their results, the maximum $\tan\beta$ value excluded is $0.88$ for $0.20< m_{H^\pm}< 0.55$~TeV. 
The corresponding searches carried out by ATLAS at $\sqrt{s}=13$~TeV and  the integrated luminosity $L = 13.2$ fb$^{-1}$ have been excluded $m_{H^+}\approx 300-900$~GeV for a very low $\tan\beta (\approx 0.5-1.7)$ region \cite{ATLAS:2016qiq}, where as for high values of $\tan\beta >44(60)$, $m_{H^+}\approx 300(366)$~GeV are excluded. 
Although, a definitive search over the $m_{H^+}-\tan\beta$ plane is a program that still has to be carried out and this belongs to the LHC experiments and future colliders.  

In this work, for our numerical analysis we restrict ourselves to the allowed regions of the $m_{H^+}-\tan\beta$ parameter space evaluated by the CMS experiments, see Fig.7 in Ref.~\cite{Sirunyan:2020hwv}. Moreover, from Ref.~\cite{Nakamura:2010zzi} we adopt other input parameters as 
$G_F = 1.16637\times10^{-5}$~GeV$^{-2}$ and
$m_t = 172.98$~GeV.
We will also evaluate the QCD coupling constant $\alpha_s$ at NLO in the $\overline{\text{MS}}$ scheme through the following relation
\begin{eqnarray}\label{alpha}
\alpha^{(n_f)}_s(\mu)=\frac{1}{b_0\ln(\mu^2/\Lambda^2)}
\Big\{1-\frac{b_1 \ln\big[\ln(\mu^2/\Lambda^2)\big]}{b_0^2\ln(\mu^2/\Lambda^2)}\Big\},
\nonumber\\
\end{eqnarray}
where,  $\Lambda$ is the QCD scale parameter. Also, $b_0$ and $b_1$ are given by 
\begin{eqnarray}
b_0=\frac{33-2n_f}{12\pi}, \quad  b_1=\frac{153-19n_f}{24\pi^2},
\end{eqnarray}
where, $n_f$ is the number of active quark flavors. In this work,  we adopt
$\Lambda_{\overline{\text{MS}}}^{(5)}=231.0$~MeV adjusted such
 that $\alpha_s^{(5)}(\mu)=0.1184$ for $\mu=m_Z=91.1876$~GeV \cite{Nakamura:2010zzi}.\\
First, we present the numerical results for the NLO decay rate $\Gamma(H^+\to t\bar{b})$ at the ZM-VFN scheme. To do this, we consider $d\Gamma_b/dx_b$ (\ref{pol1}) and integrate over $x_b (0\leq x_b\leq 1)$. Our results for various values of $m_{H^+}$ read
\begin{eqnarray}\label{rate}
\Gamma^{NLO}&=&\Gamma_0(1-0.01574),\quad \text{for}\quad m_{H^+}=200~GeV\nonumber\\
\Gamma^{NLO}&=&\Gamma_0^\prime(1-0.05396),\quad \text{for}\quad m_{H^+}=400~GeV\nonumber\\
\Gamma^{NLO}&=&\Gamma_0^{\prime\prime}(1-0.07050),\quad \text{for}\quad m_{H^+}=800~GeV.\nonumber\\
\end{eqnarray}
The decay rate at the Born level (\ref{gammatreezero}) depends on  $m_{H^+}$ and $\tan\beta$. For the tree-level decay rates in the above relations we have $\Gamma_0=0.7493\cot^2\beta$, $\Gamma_0^\prime=15.6038\cot^2\beta$ and $\Gamma_0^{\prime\prime}=42.9046\cot^2\beta$. 
From Eq.~(\ref{rate}), it is seen that the QCD corrections decrease the charged Higgs boson decay width and their amounts depend on the charged Higgs mass. Note that, for the total decay rate of process $H^+\to \bar{b} t(\to bW^+(\to l^+\nu_l))$ the above results should be multiplied by  $B(t\to bW^+)$ and $B(W^+\to l^+\nu_l)$, see Eq.~(\ref{br}). 

Now, we go back to our main aim: the evaluation of energy distribution of B-hadrons in heavy charged Higgs decays. For this purpose, we use  the factorization relation (\ref{eq:master}) where to describe the splitting  $(b, g)\rightarrow B$, from Ref.~\cite{Salajegheh:2019ach} we employ the
realistic nonperturbative $B$-hadron  FFs  determined at NLO in the ZM-VFN scheme. These FFs have been determined  through a global fit  to
electron-positron annihilation data presented by ALEPH \cite{Heister:2001jg} and OPAL
\cite{Abbiendi:2002vt} at CERN LEP1 and by SLD \cite{Abe:1999ki} at SLAC SLC. According to the approach used in \cite{Salajegheh:2019ach},  the power ansatz  $D_b(z,\mu_F^\text{ini})=Nz^\alpha(1-z)^\beta$ is adopted for the $b\to B$ splitting where  the free parameters have been determined at the initial scale $\mu_F^\text{ini}=4.5$~GeV. The fit yielded $N=2575.014$, $\alpha=15.424$, and $\beta=2.394$. The gluon FF is assumed to be zero at the initial scale $\mu_F^\text{ini}$ and generated
via the DGLAP  evolution equations \cite{dglap}.
%%%%%%%%%%%%%%%%%%%%%%%%%%%%%%%%%%%%%
\begin{figure}
	\begin{center}
		\includegraphics[width=0.7\linewidth,bb=137 42 690 690]{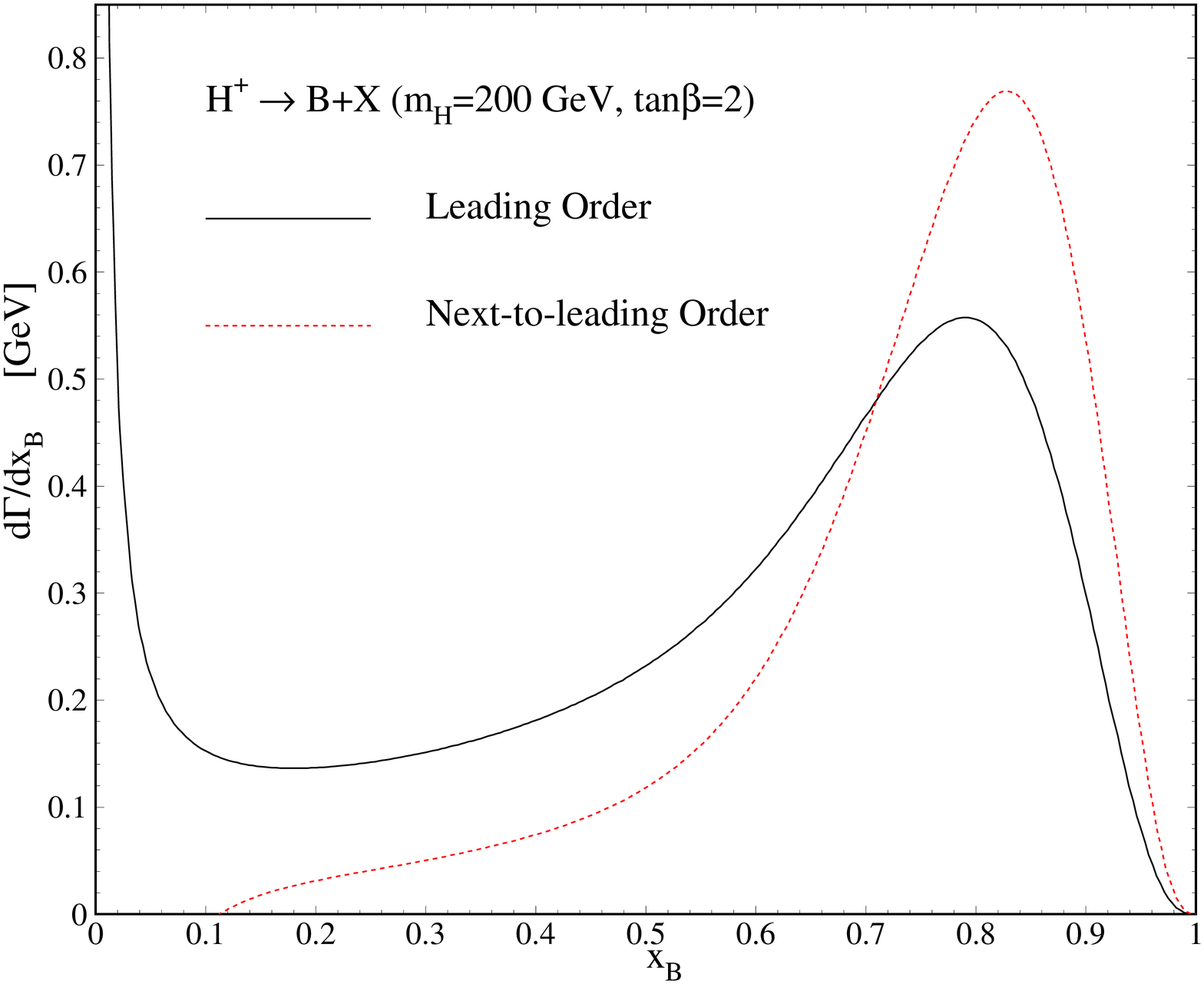}
		\caption{\label{fig2}%
			The $x_B$ spectrum  in heavy charged Higgs decay in the 2HDM. The NLO result (dashed line) is compared to the LO one (solid line) taking  $\tan\beta=2$ and  $m_{H^+}=200$~GeV. }
	\end{center}
\end{figure}
%%%%%%%%%%%%%%%%%%%%%%%%%%%%%%%%%%%%%%%
In Fig.~\ref{fig2},   our prediction for the energy spectrum of bottom-flavored hadrons is presented by plotting $d\Gamma/dx_B$ versus $x_B$. For this prediction, we have studied the size of the NLO corrections by comparing the LO (solid line) and NLO (dashed line) distributions. In order to study the importance of NLO corrections at the parton level, we evaluated the LO distribution using the same NLO $b\to B$ FF. Our results show that the NLO corrections lead to a significant enhancement of the partial decay width in the peak region and above, while these corrections decrease the size of partial decay rate in the lower-$x_B$ range. It should be noted that, the contribution of gluon splitting is appreciable only in the low-$x_B$ region.  For higher values of $x_B$,  the contribution of b-quark fragmentation dominates, as expected \cite{Kniehl:2012mn}. 
\\
In Fig.~\ref{fig1},  the dependence of $x_B$ spectrum on $\tan\beta$ is studied, taking $m_{H^+}=200$~GeV. As is seen, when the value of $\tan\beta$   increases the decay rate  decreases, because the Born rate  $\Gamma_0$ (\ref{gammatreezero}) is proportional to $\cot^2\beta$. 
\\
In Fig.~\ref{fig3},  by fixing $\tan\beta=2$ we have investigated the dependence of $x_B$ spectrum on the charged Higgs mass taking $m_{H^+}=200$ (solid line),  $m_{H^+}=400$~GeV (dashed line) and $m_{H^+}=600$~GeV (dot-dashed line). 
This figure shows that, if $m_{H^+}$  increases the size of partial decay width increases as well. Nevertheless, the peak position of $x_B$-distribution is approximately independent of the charged Higgs mass.
%%%%%%%%%%%%%%%%%%%%%%%%%%%%%%%%%%%%%%%%5
\begin{figure}
	\begin{center}
		\includegraphics[width=0.7\linewidth,bb=137 42 690 690]{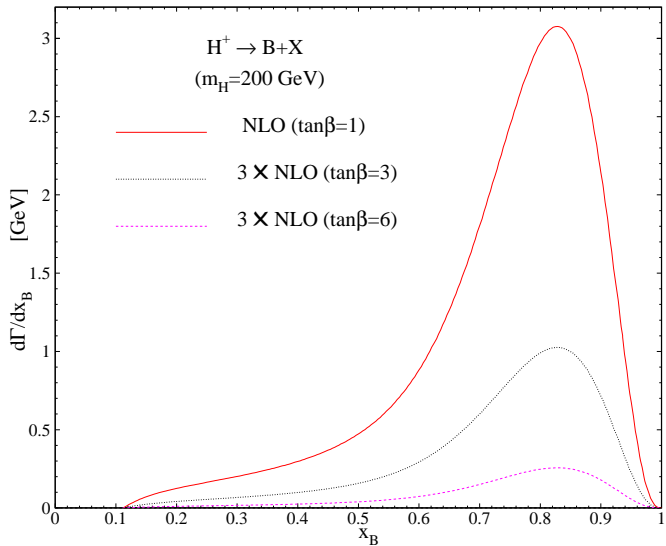}
		\caption{\label{fig1}%
			$d\Gamma(H^+\to BX)/x_B$ as a function of $x_B$   in the 2HDM considering different values of $\tan\beta=1$, $3$ and $6$. The mass of heavy charged Higgs is fixed to $m_{H^+}=200$~GeV. }
	\end{center}
\end{figure}
%%%%%%%%%%%%%%%%%%%%%%%%%%%%%%%%%%%%%%%%%%%%%%%%%
\begin{figure}
	\begin{center}
		\includegraphics[width=0.7\linewidth,bb=137 42 690 690]{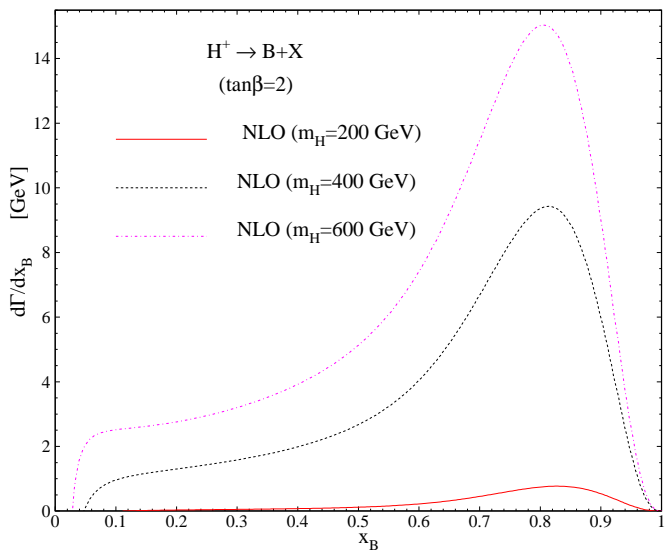}
		\caption{\label{fig3}%
			The $x_B$ spectrum in charged Higgs decay for different values of charged Higgs mass:  $m_{H^+}=200$~GeV (solid line), $m_{H^+}=400$~GeV (dashed line)  and $m_{H^+}=600$~GeV (dot-dashed line).  }
	\end{center}
\end{figure}

\section{Conclusions}
\label{sec:three}

The SM of particle physics predicts one neutral Higgs boson, whereas the Minimal Supersymmetric requires five Higgs particles, three neutral bosons 
and two charged bosons. The discovery of charged Higgs bosons would be proof of new physics beyond the SM. For this reason, searches for the charged 
Higgs bosons are strongly motivated so that in recent years it has been  a goal of many high energy colliders such as the 
CERN LHC. Searches for light charged Higgs bosons (particles lighter than the top quark) has been inconclusive and no 
evidence has been yet found. In this regard, the results reported by the CMS and ATLAS Collaborations show the large excluded  region in the MSSM $m_{H^+}-\tan\beta$ parameter space.
Therefore, it sounds that most efforts should be concentrated on  probing heavy charged Higgs bosons (heavier than the top quark). These scalar bosons are predicted to decay predominantly either to a tau and its associated neutrino ($\tau\bar{\nu}_\tau$), or to a top and a bottom quark ($t\bar{b}$). In spite of the fact that the decay channel $H^+\to t\bar{b}$  suffers from large multi-jet background, but it dominates in the heavy mass region. \\
In this work, we studied the dominant decay channel $H^+\to t\bar{b}(+g)$ followed by the hadronization process $(b,g)\to B$. At colliders, the bottom-flavored hadrons  could be identified by a displaced decay vertex associated whit charged lepton tracks. On other words, B-hadrons decay to the $J/\psi$ followed by the $J/\psi\to \mu^+\mu^-$ decays, see Ref.~\cite{Kharchilava:1999yj}. Then a muon in jet is associated to the b-flavored hadron. Furthermore, one can also explore an other way to associated the $J/\psi$ with the corresponding isolated lepton- by measuring the jet charge of identified $b$ and not requiring the tagging muon.  Therefore, at the LHC and future colliders the decay channel $H^+\rightarrow B+X$ is proposed  to search for the heavy charged Higgs bosons  and evaluating the distribution in the scaled-energy ($x_B$) of B-mesons  would be  of particular interest. This distribution is studied  by evaluating the quantity $d\Gamma/dx_B$.  
To present our phenomenological prediction of the $x_B$-distribution, we first calculated an analytic
expression for  the NLO radiative corrections to the differential   decay width $d\Gamma(H^+\to t\bar{b})/dx_a (a=b, g)$ and then
employed the nonperturbative $(b,g)\to B$ FFs, relying on their universality and scaling violations.
Our results have been  presented in the ZM-VFN scheme where the b-quark mass is ignored from the beginning. In this scheme, results are the same in both   the type-I and II 2HDM scenarios.\\
Our analysis is expected to make a contribution to the LHC searches for charged Higgs bosons. 
In fact, a comparison between the energy spectrum of B-mesons produced from charged Higgs decays at 2HDM and those from top decays at SM ($t\to B+X$) would indicate  a signal for new physics beyond the SM.
 \\
Our analysis can be also extended  to the production of hadron species other than the B-hadron, such as pions, kaons and protons, etc. This would be  possible by  using the nonperturbative $(b, g)\rightarrow \pi/K/P/D^+$ FFs presented in Refs.~\cite{Soleymaninia:2013cxa,Nejad:2015fdh,Salajegheh:2019srg,Salajegheh:2019nea}.

\end{document}